# Semiconductor superlattice diodes for detection of THz photons: The role of plasmon polariton excitation

Anatoly A. Ignatov[1,2 a)].

[1] *Institute for Physics of Microstructures, RAS, GSP-105, Nizhny Novgorod 603950, Russia*

[2] *University of Nizhny Novgorod, Nizhny Novgorod 603950, Russia*

The current (voltage) responsivity of the superlattice-based diode detectors in the terahertz frequency band which includes the region of the polar optical phonon frequencies has been analyzed theoretically. Within the framework of the equivalent circuit approach an electro-dynamical model which allows one to analyze the responsivity taking into account the plasmon polariton excitation both in the substrate and in the contact layers of the diodes has been suggested. It has been shown that the presence of the plasmon polariton excitation gives rise to strong features in the frequency dependence of the superlattice-based diodes responsivity, i.e. to the resonance dips and peaks at frequencies of hybridized plasmons and optical phonons. It has been suggested that by judicious engineering of the superlattice-based diodes it would be possible to enhance substantially their responsivity in the terahertz frequency band.

## I. INTRODUCTION

Recently, a significant progress in the development of various powerful sources of pulsed terahertz (THz) radiation has been reached. In this content one can mention the fast development of the optically pumped THz molecular lasers[1], the high-power THz radiation sources from relativistic electrons[2], the pulsed sub-terahertz and terahertz gyrotrons[3,4], and the free electron lasers (FELs)[5-7]. These sources seem to be very promising in terms of a number of applications such as THz-wave imaging, biological sciences, and pump-probe studies of dynamical properties of materials[1-7]. In this regard, a number of new concepts of THz pulsed radiation detectors having a rather high sensitivity, a short time of response, and a wide dynamic range has been put forward. Among others we can refer to traditional Schottky diode detectors[8-10], photon-drag detectors[1], field effect transistors detectors[11], THz range quantum well infrared photodetectors[12], and ultra fast graphene-based THz detectors[13].

On the other hand, superlattice-based diodes have also received considerable attention for detection of the ultra short terahertz pulses[14]. Experimentally, the interaction of the THz fields with the miniband electrons in semiconductors superlattices that show a negative differential conductance at room temperatures have been investigated in Ref. 15. A strong ac field-induced reduction of the dc current through the superlattices for the different bias voltage below and above characteristic frequency (1 THz) has been observed[16] demonstrating quasi static and dynamic regimes of the superlattice interaction with the strong terahertz fields. The experimental results have been interpreted in terms of the phase-modulated damped Bloch oscillations of electron wave packets in a sperlattice miniband[17-20].

---
[1] a) Author to whom correspondence should be addressed. Electronic mail: ign@ipm.sci-nnov.ru

Later, the superlattice-based diodes were tested with the pulsed free-electron laser (FELIX) in the Netherlands[14]. It was showed experimentally that the detector is ultra fast (200 fs) and robust with respect to the high laser power. It was demonstrated that a GaAs/AlAs superlattice detector can be used as an autocorrelator for intense THz pulses at 4.3 THz[21]. The superlattice autocorrelator was a key player in developing of a new method to measure the shape of short THz pulses by all-electronic gating[22]. Presently, superlattice-based diodes serve the purpose to detect the ELBE FEL pulses for the wavelengths above $40\ \mu\text{m}$ in Dresden (Germany)[23].

However we believe that despite the fact that since the discovery of the superlattice in 1970 by Esaki and Tsu[24] it has been several decades and numerous results on the high-frequency properties of semiconductor superlattices found their way in a number of reviews[25-27], some of the principal features of the THz photon detection with superlattices still remain untouched. This applies in particular to the question of the effect of the various electro dynamical eigenmodes excitation on the ac response of the superlattice-based diodes in the terahertz frequency band.

The present paper is devoted to a study of detection of the THz photons with superlattice-based diodes in order to provide a capability for improvements of their performance. A special emphasis is paid to clarify the role of the plasmon polariton excitation in the superlattice, substrate and in the contact layers of the superlattice-based diode in its responsivity in the terahertz (1-30 THz) frequency band. It is important to note that presently the plasmon polariton excitation start to play a key role in numerous research of nano-optical and tunable terahertz metamaterials[28-30]. We believe that methods and approaches developed in this rapidly growing field will be very useful for the study of detection of the THz photons with the superlattice-based diodes.

A strong suppression of the responsivity of the superlattice-based detectors at frequencies of the transverse polar-optic phonons and its strong increase at frequencies of the longitudinal polar-optic phonons were observed in Ref. 31. In this paper it has been demonstrated that the superlattice detector has a large frequency range (5-12 THz) and a very short time ($\tau_R < 10\ ps$) of response. Theoretically detection of the THz radiation with semiconductor superlattices at polar optical phonon frequency was investigated in Ref. 32. It was shown that that the responsivity can be suppressed at frequencies of the transversal optical phonons due to dynamic screening of the THz fields by the lattice. In contrast, the responsivity can be strongly increased at longitudinal optical phonon frequencies due to a large enhancement of the THz fields in the superlattice, i.e. due to an "anti-screening" effect. It was suggested that this phenomenon can strongly influence the performance of the superlattice-based ultra-fast detectors for THz photons. However, it should be pointed out that this paper is based on one strongly oversimplified assumption, i.e. Ref. 32 fully disregards the eigenmodes excitation both in the substrate and in the contact layers of the superlattice-based diodes.

Meanwhile, as it is well known from the Schottky diodes modelling[8-10] the charge carrier inertia, the displacement current contribution, the plasma resonance of the charge carriers, and coupling between charge carriers and the optical phonons in the substrate and in the contact layers may have a profound effect on the Schottky diodes performance. These factors to a large extent influence the frequency dependence of the Schottky diodes responsivity in the terahertz ($f > 1\ THz$) frequency band. Thus, bringing a rich experience gained in the terahertz Schottky diodes modeling to the study of the superlattice-based terahertz diodes seems to be of a considerable interest.

The paper is organized as follows. In Sec. II, we introduce the electro-dynamical modal for a study of the superlattice-based diodes that allows one to take into account the plasmon polariton excitation both in the contact layers of the superlattice and in the substrate. Basically this model relies on the equivalent circuit description of the interaction between an incident terahertz radiation with the Schottky diodes detectors[8-10].

In Sec. III, we introduce the current (voltage) responsivity of the superlattice-based diodes emphasizing the important role of both the substrate and the contact layers. We specify how the plasmon polariton excitation in the superlattice, contact layers, and in the substrate has an impact on the value and on the frequency dependence of the responsivity.

In Sec. IV, we suggest the ac/ dc electron transport model for the description of the non-linear interaction between incoming terahertz radiation with conductance electrons in a superlattice miniband. The model is based on Boltzmann equation with the relaxation-time approximation for the collision integral.

In Sec. V, we summarize the obtained results. Firstly, we demonstrate that the presence of the plasmon polariton excitation in the contact layers and in the substrate gives rise to strong resonant features in the frequency dependence of the superlattice-based diodes responsivity. Secondly, we compare our results with experiments on the responsivity measurements performed with free-electron lasers at terahertz (5-12THz) frequencies.

In Sec. VI, the conclusions are briefly summarized.

## II. ELECTRO-DYNAMICAL MODEL

Following experiments on the terahertz photons detection with two-terminal devices[8,15,21,31] let us consider a superlattice-based diode mounted in the corner-cube reflector antenna which is schematically depicted in Fig. 1(a). The incident radiation (power $P_i$, frequency $\omega$) focused to the main lobe of the antenna pattern by the curved mirror generates inside the diode intense terahertz fields (currents). In its tern, the dc current change in the external video frequency circuit takes place due to a non-linearity of the superlattice ac response.

As shown in Fig. 2(b) the superlattice-based diode is presented by multilayered hetero-junctions forming a sequence of wells and barriers with thicknesses $d_w$ and $d_b$, respectively. The superlattice mesa-structure with the period $d = d_w + d_b$, the aria $S = \pi a^2$, the mesa radius $a$, the length $L = N \times d$, and the number of the superlattice periods $N$ is imposed on a semiconductor substrate having the thickness $L_S$, the conductance electron density $n_S$, and the electron mobility $\mu_s$. In order to provide an efficient electronic contact with the metal electrodes and with the substrate the superlattice is imbedded between two contact layers having the total thickness $L_{CL}$, the conductance electron density $n_{CL}$, and the electron mobility $\mu_{cl}$.

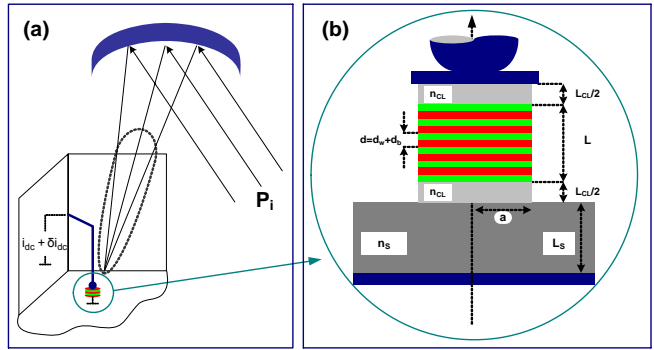

FIG. 1. (a) Schematic presentation of the superlattice-based diode mounted in the corner-cube reflector antenna. (b) Cross-section of the superlattice-based diode.

We accept a conventional equivalent circuit approach[8-10,15] for a theoretical description of the interaction of the incident terahertz radiation with the superlattice-based diode. Figure 2(a) presents schematically the antenna (having the impedance $Z_a(\omega)$) delivering the terahertz power to the superlattice-based diode described by a series connection of the superlattice impedance $Z_{sl}(\omega, V_0)$ and the series impedance $Z_s(\omega)$, where $V_0$ is the dc voltage drop on the superlattice in the operation point. Figures 2(b) and 2(c) show two kinds of the video frequency equivalent circuits normally employed for a description of the quasi-continuous incident power operation mode[8-10] and for the short-pulse incident power operation mode[1,14,15] of the detector, respectively. Finally, Fig. 2(d) presents an equivalent circuit of the superlattice-based diode taking account of the frequency-dependent capacitance of the superlattice caused by the polar optic phonons $C_0(\omega)$, as well as the conductance electrons current in the superlattice $B$. Finally, the current change in the video frequency circuits induced by the terahertz fields gives rise to the voltage drop $V_l$ on the load impedance $Z_l$ which is registered with an external electronics.

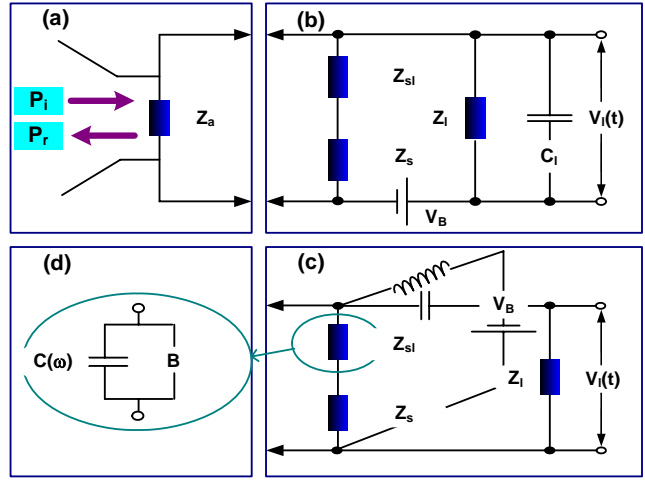

FIG. 2 (a) THz radiation coupled to a corner-cube reflector antenna, $P_i$ and $P_r$ are the incident and the reflected powers, respectively. (b) Equivalent circuit for a quasi-continuous wave detector operation mode: $Z_{sl}$-superlattice impedance, $Z_s$-series impedance, $Z_l$-load impedance, $C_l$-load capacitance, $V_B$-dc bias voltage. (c) Equivalent circuit for a short pulse detector operation mode. (d) Equivalent circuit for the superlattice: $B$-conductance electrons current, $C_0(\omega)$-frequency-dependent superlattice capacitance.

In order to take into account the plasmon polariton excitation[28,29,33] we describe the lattice dielectric function for both the wells and the barriers of the superlattice, as well as the lattice dielectric function of the substrate and of the contact layers by the Lorentzian equation

$$\varepsilon(\omega) = \varepsilon(\infty) + \left[\varepsilon(0) - \varepsilon(\infty)\right] \frac{\omega_{TO}^2}{\omega_{TO}^2 - \omega^2 + i\gamma\omega}, \qquad (1)$$

where $\varepsilon(0)$ is the low-frequency lattice dielectric constant, $\varepsilon(\infty)$ is the lattice dielectric constant at infinite frequencies $\omega \to \infty$, $\omega_{TO}$ is the frequency of the transversal polar optical phonons, and $\gamma$ is the damping constant. According to the Lyddane-Sachs-Teller equation, the longitudinal optical phonon frequency, defined by the equation $\varepsilon(\omega_{LO}) = 0$ is $\omega_{LO} = \omega_{TO}\sqrt{\varepsilon(0)/\varepsilon(\infty)}$ [33].

We calculate the superlattice effective lattice dielectric function $\varepsilon_{eff}(\omega)$ making use of the effective medium theory[33,34]

$$\frac{1}{\varepsilon_{eff}(\omega)} = \frac{1}{d_w + d_b}\left[\frac{d_w}{\varepsilon_w(\omega)} + \frac{d_b}{\varepsilon_b(\omega)}\right], \tag{2}$$

where the index $w$ refers to the wells and the index $b$ to the barriers materials.

In this case the frequency-dependent superlattice capacitance can be written as $C_0(\omega) = \varepsilon_0 \varepsilon_{eff}(\omega) S / L$, where $\varepsilon_0$ is the vacuum permittivity. Thus, the superlattice impedance can be defined as

$$Z_{sl}(\omega, V_0) = \frac{1}{G_{sl}^{ce}(\omega, V_0) + i\omega C_0(\omega)}, \tag{3}$$

where $G_{sl}^{ce}(\omega, V_0)$ is the ac superlattice electron conductance.

The total impedance of the superlattice-based diode interacting with the antenna can be presented as

$$Z_d(\omega, V_0) = Z_{sl}(\omega, V_0) + Z_s(\omega) \tag{4}$$

where $Z_s(\omega) = Z_{sub}(\omega) + Z_{cl}(\omega)$ is the series impedance, $Z_{sub}(\omega)$ is the substrate impedance, and $Z_{cl}(\omega)$ is the impedance of the contact layers.

For calculation of the substrate impedance we employ the Champlin's and Eisenstein's theory[8-10], which presents the value of $Z_{sub}(\omega)$ in the form

$$Z_{sub}(\omega) = Z_{sp}(\omega) + Z_{rad}(\omega), \tag{5}$$

where

$$Z_{sp}(\omega) = \frac{1}{2\pi a \sigma_s(\omega)} \arctan\left(\frac{L_S}{a}\right) \tag{6}$$

is the spreading impedance, and

$$Z_{rad}(\omega) = \frac{\gamma(\omega)}{2\pi \sigma_s(\omega)} \ln\left(\frac{L_S}{a}\right) \tag{7}$$

is the radiation impedance of the substrate, $\gamma(\omega) = \sqrt{i\omega\mu_0\sigma_s(\omega)}$ is the complex propagation constant of the electromagnetic waves in the substrate, $\sigma_s(\omega) = \left[\sigma_{s0}/(1+i\omega\tau_s)\right] + i\omega\varepsilon_0\varepsilon_s(\omega)$ is the complex conductivity of the substrate which takes into account both the current produced by the conduction electrons and the displacement current, $\varepsilon_s(\omega)$ is the lattice dielectric constant of the substrate, $\sigma_{s0} = en_S\mu_s$ is the static conductivity of the substrate, $\mu_s = (e/m_s)\tau_s$ is the mobility of conduction electrons in the substrate, $m_s$ is the effective mass of electrons in the substrate, $\tau_s$ is the relaxation time for electron scattering in the substrate, and $\mu_0$ is the magnetic permeability of free space.

In terms of the parallel plate capacitor approximation the complex impedance of the contact layers can be written in the form[8-10]

$$Z_{cl}(\omega) = \frac{1}{G_{cl}(\omega) + i\omega C_{cl}(\omega)}, \qquad (8)$$

where $G_{cl}(\omega) = \sigma_{cl}(\omega)S/L_{CL}$ is the complex ac conductance of the contact layers, $\sigma_{cl}(\omega) = \sigma_{cl}^{(0)}/(1+i\omega\tau_{cl})$ is the ac complex contact layers conductivity, $\sigma_{cl}^{(0)} = en_{CL}\mu_{cl}$ is the static contact layers conductivity, $\mu_{cl} = (e/m_{cl})\tau_{cl}$ is the mobility of conductance electrons in the contact layers, $m_{cl}$ is the effective mass of electrons in the contact layers, $\tau_{cl}$ is the relaxation time for electron scattering in the contact layers, $C_{cl}(\omega) = \varepsilon_0\varepsilon_{cl}(\omega)S/L_{CL}$ is the overall complex capacitance of the contact layers, and $\varepsilon_{cl}(\omega)$ is the lattice dielectric function of the contact layers.

Further, as a working example, we consider a superlattice consisting of *AlAs* barriers and *GaAs* wells. We assume that both the substrate and the contact layers are made of *GaAs* as well. For the superlattice we take the material parameters close to that ones of the bulk materials[35], i.e. $\varepsilon_w(\infty) = 10.9$, $\varepsilon_w(0) = 12.5$, $f_{TOw} = 8.01\ THz$, $\gamma_{TOw}/2\pi = 0.3\ THz$, $\varepsilon_b(\infty) = 8.5$, $\varepsilon_b(0) = 8.98$, $f_{TOb} = 10.8\ THz$, $\gamma_{TOb}/2\pi = 0.3\ THz$. According to the Lyddane-Sachs-Teller relation the frequencies of the longitudinal polar optic phonons are $f_{LOw} = 8.6\ THz$, $f_{LOb} = 11\ THz$. For the calculation of the relaxation times for electron scattering both in the substrate $\tau_s$ and in the contact layers $\tau_{cl}$ we use the Hilsum's formula for the mobility[36], i.e. $\mu = \mu_H / \left[1 + (n/n_H)^{0.4}\right]$, where $\mu_H = 10000\ cm^2/Vs$, $n_H = 10^{17}\ cm^{-3}$.

For minimization of the ohmic losses strongly doped semiconductors ($n_S = (1-4)\times 10^{18} cm^{-3}$) are routinely used as a substrate material for manufacturing of the diodes operating in the terahertz frequency band[8-10]. It is important to note that there are essential features of the electromagnetic waves propagation in the strongly doped semiconductors compare to the non-doped ones. We show in Figs. 3(a) and 3(b) the real $\text{Re}(\gamma)$ and the imaginary $\text{Im}(\gamma)$ parts of the complex propagation constant in the substrate as a function of the frequency $f$ for $n_S \to 0$ and for $n_S = 4 \times 10^{18} cm^{-3}$, respectively. It is seen from Fig. 3(a) that at $n_S \to 0$ the well-known spectrum of electromagnetic waves in a polar crystal, i.e. the bulk phonon-polariton spectrum[28,29,33] takes place. The frequency ranges $f < f_{TO}$ and $f > f_{LO}$ corresponding to the phonon polaritons propagating in a bulk polar crystal with low damping. At the same time, the frequency range $f_{TO} < f < f_{LO}$ corresponds to the region where there are no propagating electromagnetic waves i.e. to the region total reflection[28,29,33].

The situation is fundamentally different in the case of semiconductors with the high concentration of the conduction electrons. In this case the bulk plasma frequency of the conduction electrons in the substrate

$$f_{PS} = \sqrt{\frac{e^2 n_S}{\varepsilon_0 \varepsilon_s(\infty) m_S}} \qquad (9)$$

starts to play a core role separating the region of propagating waves ($f > f_{PS}$) and the region where the waves are strongly damped ($f < f_{PS}$). It is also seen from Fig. 3(b) that at $f < 1\ THz$ the real and the imaginary parts of the propagation constant are equal, i.e. $\text{Re}(\gamma) = \text{Im}(\gamma) = (\omega \mu_0 \sigma_{S0}/2)^{1/2}$. This behavior corresponds to the exponential decay of the electromagnetic waves resulting from a skin effect[8-10]. Furthermore, at frequencies close to the frequencies of the polar optical phonons $f \approx f_{TO}, f_{LO}$ the real part of the propagation constant $\text{Re}(\gamma)$ considerably exceeds the imaginary part $\text{Im}(\gamma)$. This indicates a strong attenuation of plasmon polaritons due to an absorption caused by the conduction electrons in the substrate.

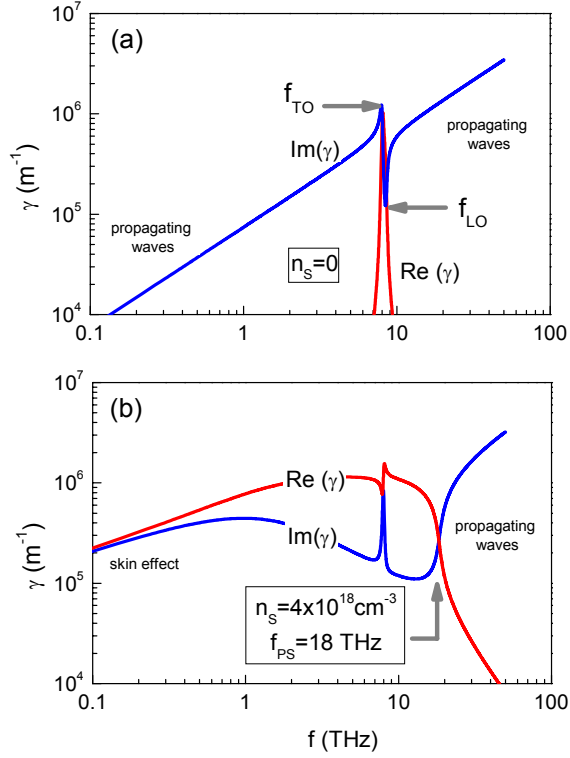

FIG. 3. The real $\text{Re}(\gamma)$ and the imaginary $\text{Im}(\gamma)$ parts of the propagating constant for electromagnetic waves in the substrate at zero (a) and at very high (b) concentration of the conduction electrons $n_S$ plotted as functions of the frequency $f$.

## III. CURRENT AND VOLTAGE RESPONSIVITIES

We suppose now that a static voltage $V_0$ is applied to the superlattice by an external bias supply and, additionally, the radiation power $P_i$ incident on the superlattice induces an alternating sinusoidal voltage with the complex amplitude $V_1$ ($|V_1| \ll V_0$)

$$V(t) = V_0 + \frac{1}{2}\left[V_1 \exp(i\omega t) + c.c.\right]. \quad (10)$$

The applied ac/dc voltage gives rise to the electric current which can be written as

$$i(t) = i_0(V_0) + \frac{1}{2}\left[G_{sl}^{ce}(\omega, V_0) V_1 \exp(i\omega t) + c.c.\right] + \delta i_{dc}(\omega, V_0), \quad (11)$$

where $i_0(V_0)$ is the static current-voltage curve of the superlattice, $G_{sl}^{ce}(\omega, V_0) = \sigma_{sl}(\omega, V_0) S/L$ is the electron conductance of the superlattice, $\sigma_{sl}(\omega, V_0)$ is the low-field conductivity of electrons in the superlattice,

$$\delta i_{dc}(\omega, V_0) = \frac{1}{4}|V_1|^2 \times \hat{i}_0''(\omega, V_0), \quad (12)$$

is the dc current change in the video-frequency part of the external circuit, and $\hat{i}_0''(\omega, V_0)$ is the generalized second derivative of the dc current-voltage curve of the superlattice. In the low-frequency limit $\omega \to 0$ the generalized second derivative of the superlattice dc current-voltage curve must reduce to the conventional definition $\hat{i}_0''(\omega, V_0) = d^2 i_0(V_0)/dV_0^2$.

The power absorbed by conductance electrons in the superlattice can be written as

$$P_{abs}^{ce} = \frac{1}{2} \operatorname{Re} G_{sl}^{ce}(\omega, V_0)|V_1|^2. \quad (13)$$

According to the video-frequency part of the equivalent circuits depicted in Figs. 2(b) and 2(c) the voltage responsivity of the superlattice detector for the quasi continuous wave operation mode $S_V^C(\omega, V_0)$ and for the pulse operation mode $S_V^P(\omega, V_0)$ can be written as[8-10]

$$S_V^C(\omega, V_0) = \frac{\delta V_l}{P_i} = S_i(\omega, V_0) \times \frac{Z_{sl}(\omega \to 0, V_0) Z_l}{Z_l + Z_s(\omega \to 0) + Z_{sl}(\omega \to 0, V_0)}, \quad (14)$$

and[1,14,15]

$$S_V^P(\omega, V_0) = \frac{\delta V_l}{P_i} = S_i(\omega, V_0) \times \frac{Z_{sl}(\omega \to 0, V_0) Z_l}{\left[Z_s(\omega \to 0) + Z_{sl}(\omega \to 0, V_0)\right]}, \quad (15)$$

where $\delta V_l$ is the voltage change on the load $Z_l$ caused by the incident THz power $P_i$,

$$S_i(\omega, V_0) = \frac{\delta i_{dc}(\omega, V_0)}{P_i} \quad (16)$$

is the current responsivity of the superlattice detector, $Z_{sl}(\omega \to 0, V_0)$ is the dc value of the differential resistance of the superlattice at the operation point, and $Z_s(\omega \to 0)$ is the dc value of the frequency-dependent series impedance $Z_s(\omega)$. It is of importance that the measured value of the voltage responsivity can be different for different types of the video-frequency circuits, i.e. $S_V^C(\omega, V_0) \neq S_V^P(\omega, V_0)$, while the current responsivity $S_i(\omega, V_0)$ is depending only on the intrinsic superlattice-based diode and on the antenna characteristics.

In order to calculate the responsivity $S_i(\omega, V_0)$ we note that according to the high-frequency part of the equivalent circuit depicted in Fig. 1(a) the squared amplitude of the ac voltage drop across the superlattice $|V_1|^2$ can be presented as

$$|V_1|^2 = 2\gamma_e(\omega)\gamma_c(\omega, V_0) \frac{|Z_{sl}(\omega, V_0)|^2}{\text{Re}[Z_d(\omega, V_0)]} \times P_i, \quad (17)$$

where $\gamma_e(\omega)$ is the electromagnetic coupling factor between the antenna an the incident beam,

$$\gamma_c(\omega, V_0) = 4\frac{\text{Re}[Z_d(\omega, V_0)] \times \text{Re}[Z_a(\omega)]}{|Z_d(\omega, V_0) + Z_a(\omega)|^2} \quad (18)$$

is the matching factor between the antenna impedance $Z_a(\omega)$ and the total superlattice diode impedance $Z_d(\omega, V_0)$[8-10].

It is important to note that according to Eq. (18) at some frequency $\omega_m$ and dc voltage $V_{0m}$ the matching factor tends to unity, i.e. $\gamma_c(\omega_m, V_{0m}) \to 1$ if the antenna impedance $Z_a(\omega_m)$ is *conjugate matched* to the total superlattice diode impedance $Z_d(\omega_m, V_{0m})$, i.e. if

$$Z_a^*(\omega_m) = Z_d(\omega_m, V_{0m}). \quad (19)$$

Making use Eqs. (16), (17) and (18), the current responsivity $S_i(\omega, V_0)$ can be written as

$$S_i(\omega, V_0) = \frac{1}{2}\gamma_e(\omega)\gamma_c(\omega, V_0) \frac{|Z_{sl}(\omega, V_0)|^2}{\text{Re}[Z_{sl}(\omega, V_0) + Z_s(\omega)]} \hat{i}_0''(V_0, \omega) =$$
$$= S_0(\omega, V_0) \times \gamma_e(\omega) \times \gamma_c(\omega, V_0) \times \gamma_a(\omega, V_0) \times \gamma_b(\omega, V_0) \quad (20)$$

where

$$S_0(\omega, V_0) = \frac{\delta i_{dc}(\omega, V_0)}{P_{abs}^{ce}}, \quad (21)$$

is the ideal current responsivity, i.e. the responsivity turning up when all incident power is absorbed by the superlattice conduction electrons,

$$\gamma_a(\omega, V_0) = \frac{P_{abs}^{sl}}{P_{abs}^d} = \frac{\text{Re}[Z_{sl}(\omega, V_0)]}{\text{Re}[Z_{sl}(\omega, V_0) + Z_s(\omega)]} \quad (22)$$

is the ratio of the power $P_{abs}^{sl}$ absorbed in the superlattice and the total power $P_{abs}^d$ absorbed in the diode, and

$$\gamma_b(\omega, V_0) = \frac{P_{abs}^{ce}}{P_{abs}^{sl}} = \frac{\operatorname{Re}\left[G_{sl}^{ce}(\omega, V_0)\right]}{\operatorname{Re}\left[G_{sl}^{ce}(\omega, V_0) + i\omega C_0(\omega)\right]} \quad (23)$$

is the ratio of the power $P_{abs}^{ce}$ absorbed in the superlattice by the conductance electrons and the total power $P_{abs}^{sl}$ absorbed in the superlattice both by the conductance electrons and by the polar optical phonons.

We note that if the real part of the superlattice free electron ac conductance is positive, i.e. if $\operatorname{Re}\left[G_{sl}^{ce}(\omega, V_0)\right] > 0$, the coefficients $\gamma_e(\omega), \gamma_c(\omega, V_0), \gamma_a(\omega, V_0), \gamma_b(\omega, V_0) < 1$ and, consequently, the absolute value of the responsivity $|S_i(\omega, V_0)|$ does not exceed the absolute value of the ideal responsivity $|S_0(\omega, V_0)|$.

Finally, equations for the squared amplitude of the ac voltage $|V_1|^2$ and the current responsivity $S_i(\omega, V_0)$ can be written as

$$|V_1|^2 = \gamma_e(\omega) \times \frac{8\operatorname{Re}[Z_a(\omega)]}{\left|1 + [Z_a(\omega) + Z_s(\omega)]/Z_{sl}(\omega, V_0)\right|^2} \times P_i =$$
$$= \gamma_e(\omega) \times \frac{8\operatorname{Re}[Z_a(\omega)]}{\left|1 + [Z_a(\omega) + Z_s(\omega)]i\omega C_0(0)\varepsilon_{tot}(\omega, V_0)/\varepsilon_{eff}(0)\right|^2} \times P_i, \quad (24)$$

$$S_i(\omega, V_0) = \frac{2\gamma_e(\omega) \times \operatorname{Re}[Z_a(\omega)] \times \hat{i}_0''(V_0, \omega)}{\left|1 + [Z_a(\omega) + Z_s(\omega)]i\omega C_0(0)\varepsilon_{tot}(\omega, V_0)/\varepsilon_{eff}(0)\right|^2}, \quad (25)$$

where

$$\varepsilon_{tot}(\omega, V_0) = \varepsilon_{eff}(\omega) + \frac{\sigma_{sl}(V_0, \omega)}{i\omega\varepsilon_0} \quad (26)$$

is the total dielectric function of the superlattice

It should be noted that in the absence of the series resistance $Z_s(\omega) \equiv 0$ and in a particular case when the antenna impedance $Z_a(\omega)$ is purely real and frequency independent, the expressions for the ac voltage amplitude $|V_1|^2$ and for the current responsivity $S_i(\omega, V_0)$ given by Eqs. (24) and (25) reduce to that ones found in Ref. 32.

**IV. ELECTRON TRANSPORT MODEL**

For description of the ac/dc transport of conductance electrons in a superlattice we employ the semi-classical theory based on Boltzmann equation with the relaxation-time approximation for the collision integral[32,37]. Within the framework of this theory the static current-voltage curve of the superlattice can be written as

$$i_0(V_0) = i_p \frac{2(V_0/V_p)}{1+(V_0/V_p)^2}. \tag{27}$$

Equation (27) exhibits the Esaki-Tsu negative differential conductance effect[24], i.e. $di_0(V_0)/dV_0 < 0$ at $V_0 > V_p$, where

$$i_p = en \times S \times v_0 \times \frac{1}{2} \frac{I_1(\Delta/2kT)}{I_0(\Delta/2kT)} \tag{28}$$

is the maximum (peak) current at $V_0 = V_p$,

$$V_p = N \times \hbar \nu / e \tag{29}$$

is the peak voltage, i.e. the voltage at which the Esaki-Tsu negative differential conductance sets in, $\nu = 1/\tau$ is the electron scattering frequency, and $\tau$ is the electron scattering relaxation time, $j_p = i_p/S$ is the peak current density, $e$ is the electron charge, $n$ is the conductance electron density in the superlattice, $\Delta$ is the superlattice miniband width, $v_0 = \Delta d / 2\hbar$ is the maximum group velocity of electrons along the superlattice axis, $I_0(x)$, $I_1(x)$ are the modified Bessel functions, $k$ is the Boltsmann's constant, $T$ is the lattice temperature at thermal equilibrium.

At $kT \ll \Delta$ Eq. (28) recovers the original Esaki-Tsu formula for the peak current[24] $i_p = enS \times v_p$, where $v_p = v_0/2$ is the maximum (peak) drift velocity of the conductance electrons in the superlattice. The factor $I_1(\Delta/2kT)/I_0(\Delta/2kT)$ which reduces the peak current $i_p$ at elevated temperatures describes the thermal saturation of electron transport in a superlattice miniband. This effect was experimentally observed through the measurements of the Drude conductivity of the superlattice[38]. A good agreement between the theory and experimental results was obtained laying a foundation for the semi-classical description of the electron transport in a superlattice miniband.

For the electron conductance of the superlattice $G_{sl}^{ce}(\omega, V_0)$ the relaxation time approximation model delivers[32,37]

$$G_{sl}^{ce}(\omega, V_0) = G_0 \frac{1 + i\omega\tau - (V_0/V_p)^2}{\left[1 + (V_0/V_p)^2\right]\left[(V_0/V_p)^2 + (1 + i\omega\tau)^2\right]}, \quad (30)$$

where $G_0 = 2i_p/V_p$ is the low-field dc conductance of the superlattice.

At zero frequencies $\omega \to 0$ Eq. (30) reduces to the static differential conductance of the superlattice, i.e. $G_{sl}^{ce}(\omega \to 0, V_0) \to di_0(V_0)/dV_0 = G_0\left[1 - (V_0/V_p)^2\right]/\left[1 + (V_0/V_p)^2\right]^2$ pointing at the negative differential conductance effect, i.e. $G_{sl}^{ce}(\omega \to 0, V_0) < 0$ at $V_0 > V_p$. While at zero dc voltage $V_0 \to 0$ Eq. (30) resolves into the conventional Drude formula $G_{sl}^{ce}(\omega, V_0 \to 0) = G_0/(1 + i\omega\tau)$ that has been used for an analysis of the experimental results in Ref. 38.

It is remarkable to note that according to the semi classical theory[32,37] both the real part of the superlattice conductance $\text{Re}\left[G_{sl}^{ce}(\omega, V_0)\right]$ and the generalized second derivative $\hat{i}_0''(\omega, V_0)$ can be presented in a finite difference form, i.e. as

$$\text{Re}\, G_{sl}^{ce}(\omega, V_0) = \frac{1}{2} \times \frac{e}{N\hbar\omega} \times \left[i_0(V_0 + N\hbar\omega/e) - i_0(V_0 - N\hbar\omega/e)\right], \quad (31)$$

and

$$\hat{i}_0''(\omega, V_0) = \frac{1}{(N\hbar\omega/e)^2} \times \left[i_0(V_0 + N\hbar\omega/e) - 2i_0(V_0) + i_0(V_0 - N\hbar\omega/e)\right], \quad (32)$$

respectively. This presentation illustrates the intimate relationship between the absorption (emission) processes in a superlattice described by the real part of the ac differential conductance $\text{Re}\left[G_{sl}^{ce}(\omega, V_0)\right]$ and the non-linear dc current change caused by absorbed photons described by the generalized second derivative $\hat{i}_0''(\omega, V_0)$[32,37].

Using Eqs. (13), (21), (31) and (32) for the ideal current responsivity $S_0(\omega, V_0)$ one can get[32,37]

$$S_0(\omega, V_0) = -\frac{1}{V_p} \times \frac{(V_0/V_p)\left[3 + (\omega\tau)^2 - (V_0/V_p)^2\right]}{\left[1 + (V_0/V_p)^2\right]\left[1 + (\omega\tau)^2 - (V_0/V_p)^2\right]}. \quad (33)$$

In the high-frequency limit $\omega\tau \gg 1$, the ideal responsivity $S_0(\omega,V_0)$ tends to the frequency-independent value $S_0(\omega,V_0) = -(1/V_p) \times (V_0/V_p)/\left[1+(V_0/V_p)^2\right]$ which reproduces the dc Esaki-Tsu current-voltage characteristics[32-37]. It is seen from this equation that in the high-frequency limit the maximum value of the responsivity $S_0(\omega,V_0)$ is achieved at $V_0 = V_p$. Further we will use this value of the dc voltage $V_0$ for our analysis of the responsivity $S_i(\omega,V_0)$ taking into account of the eigenmodes excitations both in the substrate and in the contact layers.

**V. PLASMON POLARITON EXCITATION**

In this chapter we demonstrate the role of the plasmon polariton excitations in the frequency dependence of the responsivity $S_i(\omega,V_0)$ by the example of the broadband semiconductor superlattice detector for terahertz radiation described in Ref. 31. A superlattice used in this experiments consists of 100 periods of 14 monolayers of $GaAs$ and 3 monolayers of $AlAs$ (the superlattice period $d \approx 4.81\ nm$, the superlattice length $L = Nd = 0.48\ \mu m$). It was homogeneously doped with silicon and had the conductance electrons density $n \approx 8 \times 10^{16}\ cm^{-3}$.

In order to provide low-resistance electronic contacts with the metal electrodes and the substrate the superlattice was embedded between two 300 nm $GaAs$ thick contact layers (doping $n_{CL} \approx (1-2) \times 10^{18}\ cm^{-3}$). From the band-gap offset ($\approx 1\ eV$) between $GaAs$ wells and $AlAs$ barriers, the width of lowest miniband $\Delta = 70\ meV$, and the gap to the higher miniband $\Delta_G = 415\ meV$ have been estimated in the envelope-function approximation[39]. A small-area mesa (1 $\mu m$ diameter) was prepared for irradiation by the terahertz power produced by the free-electron laser (FELIX) in Nieuwegen, the Netherlands, delivering ultra short (duration 30 $ps$) micro pulses.

Terahertz radiation was coupled to the small-aria superlattice device via a corner-cube antenna system (see Fig.1). The impedance of the antenna system $Z_a(\omega)$ and the electromagnetic coupling factor between the antenna an the incident beam $\gamma_e(\omega)$ have been considered as frequency independent and estimated as $Z_a = 100\ \Omega$ and $\gamma_e = 2 \times 10^{-3}$, respectively.

The dc current-voltage characteristics measured in Ref. 31 for this sample showed the Esaki-Tsu negative differential conductance at $V_0 > V_p = 0.7\ V$ with the peak-current density $j_p = 100\ kA/cm^2$. For the superlattice parameters given above the measured value of the peak current $j_p$ is in a good agreement with the value of the peak current calculated from Eq. (28). Basing on the measured value of the peak voltage $V_p$ and on the number of the superlattice periods $N = 100$, the value of the conductance electrons scattering frequency can be estimated as $\nu \simeq 10^{13}\ s^{-1}$ ($\tau \simeq 10^{-13}\ s$). Below we will use the superlattice parameters quoted above as a working example for the numerical calculation of the responsivity $S_i(\omega, V_0)$.

Figure 4 (a) demonstrates the absolute value of the current responsivity $|S_i(\omega, V_0)|$ calculated as a function of the frequency $f$ at $V_0 = V_p$ for the conductance electrons concentration in the substrate $n_S = 2 \times 10^{18}\ cm^{-3}$, and for the substrate thickness $L_S = 100\ \mu m$. Three different values of the conductance electrons concentration in the contact layers have been taken for the calculations, i.e. $n_{CL} = 2 \times 10^{18}\ cm^{-3}$, $n_{CL} = 1.5 \times 10^{18}\ cm^{-3}$, and $n_{CL} = 1 \times 10^{18}\ cm^{-3}$ (curves 2, 3, and 4, respectively). The curve 1 shows the result of Ref. 31 that has been obtained disregarding the role of the substrate and the contact layers, i.e. at $n_S = \infty$ and $n_{CL} = \infty$. The weighted value of the ideal responsivity $|\gamma_e S_0(\omega, V_0)|$ is shown for comparison.

It is seen from Fig. 4 (a) that in the absence of the substrate and contact layers the responsivity $|S_i(\omega, V_0)|$ demonstrates two distinct dips at frequencies of the transversal optical phonons in the wells and in the barriers, i.e. at $f = f_{TOw}$ and $f = f_{TOb}$, respectively. Obviously, the dip at $f = f_{TOb}$ is much less pronounced than the dip at $f = f_{TOw}$ due to the low volume fraction of $AlAs$ barriers in the superlattice in comparison with the volume fraction of the $GaAs$ barriers. The curve 2 in Fig. 4 (a) demonstrates the role of the substrates and the contact layers in the value and frequency dependence of the responsivity $|S_i(\omega, V_0)|$. Firstly, it is seen that for all frequencies the presence of the substrate and the contact layers essentially lessens the value of the responsivity due to a supplementing dissipation of the terahertz power. Secondly, in addition to the dips at the transversal optical phonons frequencies $f = f_{TOw}$ and $f = f_{TOb}$ a huge dip at plasma frequencies of the conductance electrons in the substrate and the contact layers $f = f_{PS}, f_{PCL}^{(2)}$ (which are coincident for the curve 2) takes place.

With a decrease of the conduction electron density, i.e. at $n_{CL} = 1.5 \times 10^{18} cm^{-3}$ the additional dip on the curve $|S_i(f)|$ corresponding to the frequency $f_{PCL}^{(3)} < f_{PS}$ occurs but the dip at the frequency $f_{PS}$ stays up (see curve 3). At even lower densities of the conduction electrons, i.e. at $n_{CL} = 1 \times 10^{18} cm^{-3}$, position of the dip associated with the excitations of plasma oscillations in the contact layers coincides with the frequency of the transversal optical phonon in the barriers $f_{TOb}$. Moreover, near the frequencies of the transversal optical phonons $f \approx f_{TOw}, f_{TOb}$ the fine structure in the frequency dependence of the responsivity $|S_i(f)|$ occurs. This curious feature can be attributed to the effect of hybridization of plasma and optical phonon modes in the superlattice diode or, to the plasmon polariton excitation[28, 29].

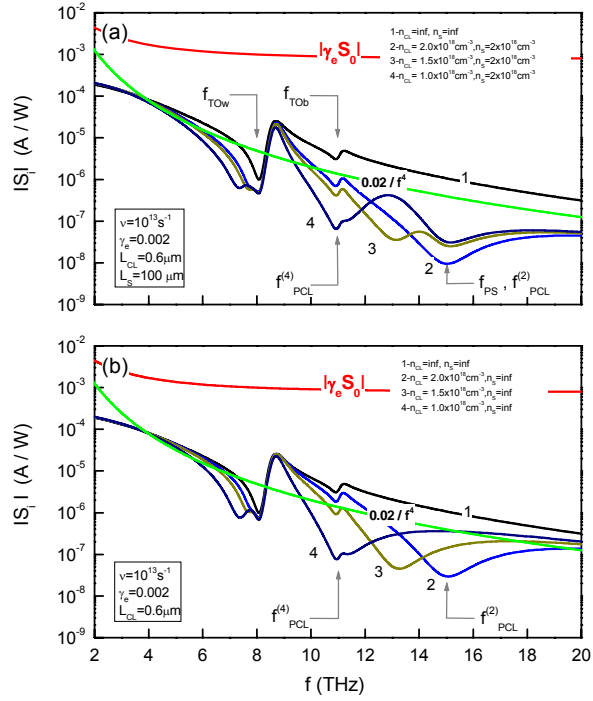

FIG. 4. Current responsivity of the superlattice diode $|S_i|$ calculated as a function of the frequency $f$ for the total thickness of the contact layers $L_{CL} = 0.6 \mu m$, for the electron scattering frequency $\nu = 10^{13} s^{-1}$, and for the conductance electron densities in the substrate $n_S = 2 \times 10^{18} cm^{-3}$ (a), and for $n_S = \infty$ (b). Curve 1 presents the result of Ref. 32 obtained disregarding the eigenmodes excitation both in the substrate and in the contact layers ($n_{CL} = \infty$, $n_S = \infty$). Different values of the conductance electron densities in the contact layers, i.e. $n_{CL} = 2 \times 10^{18} cm^{-3}$ (curves 2) $n_{CL} = 1.5 \times 10^{18} cm^{-3}$ (curves 3), and $n_{CL} = 1 \times 10^{18} cm^{-3}$ (curves 4), have been chosen for calculations. The normalized ideal current responsivity $\gamma_e |S_0|$ is also shown for comparison. The electromagnetic coupling factor between the antenna an the incident beam is taken as $\gamma_e = 0.002$.

Figure 4 (b) shows the calculated responsivity $|S_i|$ for the same values of $n_{CL}$ as in Fig. 12 (a) and for the case of the removed substrate, i.e. for $n_{CL} = \infty$. It is seen from this figure that the obtained curves are very close to the curves shown in Fig. 12 (a) apart from the absence of the resonance dip at the plasma frequency of electrons in the substrate $f = f_{PS}$. Just as in the previous figure with decreasing of the electron density in the contact layers $n_{CL}$ the resonant dip at the plasma frequency $f_{PCL}$ goes down. A general tendency of the responsivity fall with increasing of the frequency is well described by the power law $S_i \propto 1/f^4$ in both cases. Moreover, in the frequency region close to the optical phonon frequencies $f \approx f_{TOw}, f_{TOb}$ the values of the responsivities shown in Figs. 4(a) and 4(b) are very close.

Experimentally, the frequency dependence of the current responsivity of the superlattice-based terahertz photon detectors was investigated in Ref. 31 by use of the free electron laser in a wide frequency range (5-12 THz). The responsivity showed strong minima close to the frequencies of the transversal optical phonons both in the barriers and wells. Surprisingly, the resonant dip close to the frequency $f_{TOb}$ appeared to be much more pronounced than the resonant dip close to the frequency $f_{TOw}$. It is important to note that with an allowance for the small value of the electromagnetic coupling factor estimated in Ref. 31 as $\gamma_e \approx 0.002$ the measured values of the responsivity and its frequency dependence are very close to the responsivity values presented in Fig. 12 (a) and (b). Possibly, the huge dip at the frequency $f_{TOb}$ observed in Ref. 31 can be attributed to the effect of plasmon polariton excitation described above. However, a detailed comparison between our calculations and experimental resalts is difficult because of the low frequency resolution of the responsivity measurements presented in Ref. 31.

Finally, we would like to note that the semi-classical approach employed in the present paper can be justified if the characteristic length of the electron spatial localization $X = \Delta/eE_0$ in the dc electric field $E_0 = E_p = \hbar\nu/ed$ is smaller than the total length of the superlattice $L = N \times d$. For the total number of the superlattice periods $N$ this condition can be written as $N > N_0 = \Delta/\hbar\nu$. For the typical value of the characteristic scattering frequency $\nu = 10^{13} s^{-1}$ taken in our calculations the characteristic number of the superlattice periods $N_0 \approx 10$ being much less than the total number of periods $N = 100$.

On the other hand, our semi-classical approach does not take into account the inter-miniband transitions in the terahertz field irradiated superlattices. Consequently, the following condition for the frequency of the incident radiation $f < f_G = \Delta_G / 2\pi\hbar$ should be satisfied. For the sample described in the present paper we obtain $f_G \approx 100\ THz$. Therefore, the actual frequency band where the resonant features of the superlattice diode response caused by the plasmon polariton excitations have been discussed is situated well below the frequency of the inter-miniband transitions.

**VI. CONCLUSIONS**

In conclusion, we investigate the current (voltage) responsivity of the superlattice-based detectors in the terahertz frequency band. Making use the equivalent circuit approach we formulate the electro-dynamical model that allows one to analyze the superlattice response taking into account the plasmon polariton excitations in the contact layers and in the substrate of the superlattice diode.

We demonstrate firstly that the presence of the plasmon polariton excitations gives rise to strong features in the frequency dependence of the superlattice diodes responsivity, i.e. to the resonance dips at the frequencies of plasmons and optical phonons in the superlattice, substrate and contact layers. Secondly we show that a specific fine structure in the dependence of the responsivity as a function of the frequency can occur if the plasma frequency of the conduction electrons either in substrate or in the contact layers is coincident with the optical phonons frequencies due to the hybridization of the plasmon and polariton oscillations.

We suggest that our study may be suitable for the design and optimization of the superlattice-based detectors in the wide frequency band which includes the plasma and the polar optical phonons frequencies. We believe that by judicious engineering of the superlattice-based diode geometry and the superlattice samples parameters it would be possible to enhance substantially the responsivity of the superlattice-based detectors operating at terahertz frequencies.

**VII. ACKNOWLEDGMENTS**

The author would like to thank M. Bakunov, A. Eisfeld, M. Helm, D. Paveliev, and S. Winnerl for helpful discussions.

The support from the Program for the Basic Research of the Russian Academy of Sciences and from the Max Planck Society for the Advancement of Science (Germany) is gratefully acknowledged.